# Electric Field-Induced Second Order Nonlinear Optical Effects in Silicon Waveguides

E. Timurdogan[1], Christopher V. Poulton[1] and M. R. Watts[1]

[1] *Research Laboratory of Electronics, Massachusetts Institute of Technology, Cambridge, Massachusetts 02139, USA*

**The demand for nonlinear effects within a silicon platform to support photonic circuits requiring phase-only modulation, frequency doubling, and/or difference frequency generation, is becoming increasingly clear. However, the symmetry of the silicon crystal inhibits second order optical nonlinear susceptibility, $\chi^{(2)}$. Here, we show that the crystalline symmetry is broken when a DC field is present, inducing a $\chi^{(2)}$ in a silicon waveguide that is proportional to the large $\chi^{(3)}$ of silicon. First, Mach-Zehnder interferometers using the DC Kerr effect optical phase shifters in silicon ridge waveguides with p-i-n junctions are demonstrated with a $V_\pi L$ of 2.4Vcm in telecom bands ($\lambda_\omega$=1.58μm) without requiring to dope the silicon core. Second, the pump and second harmonic modes in silicon ridge waveguides are quasi-phase matched when the magnitude, spatial distribution of the DC field and $\chi^{(2)}$ are controlled with p-i-n junctions. Using these waveguides, second harmonic generation at multiple pump wavelengths are observed with a maximum efficiency of $P_{2\omega}/P\omega^2$=12%/W at $\lambda_\omega$=2.29μm in a 1mm long waveguide. This corresponds to a field-induced $\chi^{(2)}$=41pm/V, comparable to non-centrosymmetric media (LiNbO$_3$, GaAs, GaN). The field-induced nonlinear silicon photonics will lead to a new class of CMOS compatible integrated devices spanning from near to mid infrared spectrum.**

Second order nonlinear optical effects are not present in the complementary metal-oxide-semiconductor (CMOS) compatible materials (e.g. Si, SiN, SiO$_2$ and Ge). This stems from the fact that these materials are centro-symmetric and second order nonlinear susceptibility, $\chi^{(2)}$, in a centro-symmetric media is inhibited in the electric-dipole approximation [1]. This provides a significant challenge for second order electro-optic effects or second harmonic generation (SHG) in the CMOS compatible platform. Recently, $\chi^{(2)}$ in silicon was realized through material engineering. A SiN stressor layer was deposited on a silicon waveguide to induce a large stress gradient, breaking the centro-symmetry. SHG in this waveguide was demonstrated with an efficiency of $P_{2\omega}/P_\omega^2=10^{-5}\%/W$ and an estimated $\chi^{(2)}\sim44pm/V$ [2]. The low efficiency is attributed to the phase mismatch between the pump and signal wavelengths. Further, SHG in a SiN resonator was also demonstrated with an efficiency of $P_{2\omega}/P_\omega^2=0.1\%/W$ [3]. The efficiency was low due to the relatively small $\chi^{(2)}<0.04pm/V$ that originated from the SiN surface. Though these methods have realized SHG in a CMOS compatible platform, the efficiency was low compared to SHG with non-centro-symmetric $\chi^{(2)}$ media in other integrated platforms (e.g. LiNbO$_3$, $P_{2\omega}/P\omega^2=204\%/W$ [4], GaAs, $P_{2\omega}/P\omega^2=6.6\%/W$ [5], GaN, $P_{2\omega}/P\omega^2=0.02\%/W$ [6]). High SHG efficiency is required to stabilize octave-spanning combs for precise frequency references [7]. The demonstrations in other integrated platforms was possible due to large $\chi^{(2)}$, high modal overlap and phase matching between the pump and signal modes.

When an electric field is applied to a centro-symmetric material, the dipole moments orient in the direction of this field, inducing an asymmetry in the harmonic oscillator model and breaking the crystalline symmetry [8,9]. The third order nonlinear susceptibility, $\chi^{(3)}$, is converted to $\chi^{(2)}$ through this process. The field-induced second order nonlinear effect in the form of SHG was observed in silicon metal-oxide-semiconductor (MOS) transistors or Schottky contacts [9]. A pump laser was focused to the gate of the MOS transistors or Schottky contacts and the reflected second harmonic signal below the silicon transparency region (i.e. λ<1.07μm) was measured as a function of applied DC bias. The generated second harmonic signal was proportional to the electric field. However, SHG efficiency was not determined and the efficiency is expected to be small due to the nanometer scale interaction length and large optical losses below silicon transparency. This effect is strong in silicon because silicon exhibits a large $\chi^{(3)}$ compared to other common CMOS compatible materials such as SiN and $SiO_2$ [10]. In fact, field induced nonlinear optical effects based on high order susceptibilities scale with an applied electric field [1,9] and the upper limit of the field-induced $\chi^{(2)}$ is imposed by the silicon breakdown field, $E_{DC}^{x}$~40V/μm [11]. The field-induced $\chi^{(2)}$ can also act on the signal to perturb the electric permittivity. This is referred to as the DC Kerr effect or quadratic electro-optic Kerr effect [10,12], an inherently phase matched process.

Here, we demonstrate the electric field-induced DC Kerr effect and second harmonic generation (EFISHG) in integrated silicon ridge waveguides. The waveguides were implanted with ions to form compact p-i-n junctions, which concentrate electric fields and utilize the large $\chi^{(3)}$ of silicon. The DC Kerr effect is demonstrated in a silicon Mach-Zehnder interferometer with a $V_\pi L$ of 2.4Vcm. For efficient EFISHG in silicon, the fundamental pump and signal modes are quasi-phase matched with periodically formed p-i-n junctions. In this configuration, a SHG efficiency of $P_{2\omega}/P\omega^2$=12%/W at $\lambda_\omega=2\lambda_{2\omega}$=2.29µm was measured, the most efficient CMOS compatible SHG to date. This corresponds to an effective field-induced $\bar{\chi}^{(2)}$ of 41pm/V at an applied DC bias of -21V. When the waveguide width and the spatial period were changed, SHG was recorded at multiple wavelengths spanning from $\lambda_\omega$=2.16µm to $\lambda_\omega$=2.42µm. The signal power scaled quadratically with pump power and proportionally with the applied reverse bias that verifies the nonlinear origin of the EFISHG. The SHG spectrum follows a sinc$^2$-like envelope, showing good agreement with theory.

The DC Kerr effect is an inherently phase matched four-wave mixing process, involving two fundamental fields ($e_\omega$) and two DC fields ($E_{DC}$) in the form of $\chi_{ijkl}^{(3)}(-\omega;\omega,0,0)$ [10]. EFISHG is a four-wave mixing process that involves two pump fields ($e_\omega$), an output harmonic signal field ($e_{2\omega}$) and a DC field ($E_{DC}$) in the form of $\chi_{ijkl}^{(3)}(-2\omega,\omega,\omega,0)$ [10]. For an applied DC and optical fields along the $x$ direction, $E_{DC}^x$ and $e_\omega^x$, the nonlinear displacement current of interest of isotropic silicon [1,10] is

$$D = \varepsilon_0 \left[ \underbrace{\varepsilon_{Si} e_\omega^x \cos(\omega t)}_{\text{Linear Polarization}} + \underbrace{12\chi_{xxxx}^{(3)} E_{DC}^{x^2} e_\omega^x \cos(\omega t)}_{\text{Electro-optic DC Kerr Effect}} + \underbrace{3\chi_{xxxx}^{(3)} E_{DC}^x e_\omega^{x^2} \cos(2\omega t)}_{\text{Field Induced Second Harmonic (EFISHG)}} \right] \quad (1)$$

where $\varepsilon_0$ and $\varepsilon_{Si}$ are the vacuum and relative silicon permittivities, respectively. The EFISHG and DC Kerr effects induce a relative permittivity related to the $\chi_{xxxx}^{(3)}$ tensor element and the applied DC field.

A silicon ridge waveguide with an embedded p-i-n junction, shown in Figure 1.a, was designed to demonstrate the DC Kerr effect. The core of the silicon waveguide is 500nm wide with an intrinsic region, $w_i$. The $w_i$ was simulated using Synopsys's Sentaurus software suite, fitting to $w_i = 200 + 58\sqrt{V_{DC} + 0.5}$ in nanometers. This was chosen to minimize the relative electro-refractive permittivity change due to the plasma-dispersion effect [11]. The fundamental TE mode was selected for maximal confinement and propagation was chosen to be in the $z$ direction. The optical mode profile, $e_\omega^x$, shown in Figure 1.b, was simulated using a finite difference mode solver [13]. The $x$, $y$ and $z$ directions are aligned with the (0 1 0), (0 0 1) and (1 0 0) crystalline axes of the silicon wafer. The generated electric field is aligned with $x$ direction (i.e. $E_{DC}^y = E_{DC}^z = 0$) for utilizing the diagonal (i.e. largest) tensor elements in the third order nonlinear susceptibility of silicon, $\chi_{xxxx}^{(3)} = 2.45 \times 10^{-19}\,\text{m}^2\text{V}^{-2}$ at $\lambda \sim 1.55\mu\text{m}$ [14]. This is realized with the lateral p-i-n junction. Using Eq. (1), the DC Kerr relative permittivity is expressed as $\Delta\varepsilon_{d.c.\,\text{Kerr}} = 12\chi_{xxxx}^{(3)} E_{DC}^{x^2}$, where the electric field is $E_{DC}^x = V_{DC}/w_i$ and $V_{DC}$ is the reverse bias. Assuming a small perturbation to the refractive index, the index perturbation was approximated with $\Delta n_{DC\,\text{Kerr}} = \Delta\varepsilon_{DC\,\text{Kerr}}/2\sqrt{\varepsilon_{Si}}$ [15]. The overlap integral between the intrinsic region and the optical mode was used to determine effective DC Kerr index perturbation, $\Delta n_{eff} = \int_v \Delta n_{d.c.\,\text{Kerr}} e_\omega^x e_\omega^{x*} dv$, where the optical mode ($e_\omega^x$) was normalized using $\int_v e_\omega^x e_\omega^{x*} dv = 1$. The DC Kerr index perturbation was simulated as a function of bias voltage, shown in Figure 1.e. Since the intrinsic silicon width slightly changes with the applied voltage, the plasma dispersion effect was also simulated using the method in [15] and shown in Figure 1.e.

The designed silicon ridge waveguides are placed in Mach-Zehnder interferometers (MZIs) as optical phase shifters for characterizing DC Kerr relative permittivity (Figure 1.c). MZIs with 3mm and 4mm long phase shifters were fabricated on a 300mm silicon-on-insulator (SOI) wafer (see Methods-Fabrication). A continuous-wave (CW) laser at λ~1580nm was coupled through a single mode fiber (SMF-28) to an inverse silicon taper. The linearly polarized output of the SMF-28 and the fundamental TE mode of the on-chip waveguide were aligned using a fiber polarization controller. The laser power is split into two arms with a broadband silicon 3dB coupler. The optical path difference between the Mach-Zehnder (MZ) arms was minimized with a silicon heater in one of the MZ arms and the MZ arms were interfered using the silicon 3dB coupler. Then, the bar and cross ports of the MZIs were recorded as a function of applied voltages to a single arm of the MZIs (Figure 1.d), demonstrating a $V_\pi L$ of 2.4Vcm. The other arm of the MZI was also doped for minimizing loss difference between the MZ arms. The insertion losses due to the fiber couplers were subtracted. The power difference between the cross and bar ports were normalized and fitted with $\cos(2\pi\Delta n(V_{DC})L/\lambda)$, where $L$ is the Mach-Zehnder arm length and $\Delta n(V_{DC})$ is the induced refractive index as a function of applied voltage, plotted in Figure 1.e. This shows good agreement with the plasma dispersion effect for small electric fields (i.e. voltages) and the DC Kerr effect for large electric fields, confirming the existence of field induced nonlinear effects in silicon.

For EFISHG, the *x* polarized pump and signal modes ($e_\omega^x, e_{2\omega}^x$) in a silicon ridge waveguide were both chosen to be fundamental TE modes, shown in Figure 2.a-b.

Fundamental modes achieve maximal confinement in the silicon core and a large overlap between the pump and second harmonic signals. The generated DC electric field is applied along *x* direction with a lateral p-i-n junction (Figure 2.c) to align with the fundamental modes. The silicon core is left intrinsic to minimize free-carrier losses and generate a uniform DC electric field. In this case, the EFISHG permittivity is related to the induced second order susceptibility using Eq. (1), $\chi_{xxx}^{(2)} = 3\chi_{xxxx}^{(3)} E_{DC}^x$ [10]. Signal and pump wavelengths at $\lambda_{2\omega}$~1.145μm and $\lambda_\omega$~2.29μm were chosen for the design. The signal wavelength was selected within the transparent silicon region to minimize absorption. The model in [16] and the third order nonlinearities in [14, 16-22] were used to estimate the tensor element to be $\chi_{xxxx}^{(3)} = (6 \pm 3.5) \times 10^{-19} \text{m}^2/\text{V}^2$ at $\lambda_\omega$~2.29μm. This uncertainty is due to large differences in reported $\chi^{(3)}$ of silicon around $\lambda_\omega$~2.29μm [16-22]. Then, the bulk second order nonlinearity within the silicon waveguide can be as large as $\chi_{xxx}^{(2)} = 3\chi_{xxxx}^{(3)} E_{DC}^x = 72 \pm 42$ pm/V for an applied field that is equal to the silicon breakdown field, $E_{DC}^x$ =40V/μm. The effective second order nonlinear susceptibility $\bar{\chi}_{xxx}^{(2)}$ that acts upon the pump and signal modes, is determined using the overlap integral over the intrinsic silicon area, $v_i$, and waveguide area, $v_0$, $\bar{\chi}_{xxx}^{(2)} = 3\chi_{xxxx}^{(3)} \sqrt{v_i} \left| \int_{v_0} e_{2\omega}^{x*} e_\omega^x e_\omega^{x*} E_{DC}^x \, dv \right|$. The pump and signal modes ($e_\omega^x, e_{2\omega}^x$) were normalized using $\int_v e_{\omega,2\omega}^x e_{\omega,2\omega}^{x*} dv = 1$. The core of the silicon waveguide was chosen to be 800nm wide for maximizing the overlap integral between the fundamental TE pump and signal modes while minimizing the required electrical voltage for generating large electric fields. The electric field within the silicon core waveguide was simulated using Synopsys's Sentaurus software suite. When a reverse bias of 21V is applied to the junction, the electric field is quite uniform inside the silicon core with $E_{DC}^{x}$ =25V/μm and the effective second order nonlinearity was simulated to be $\bar{\chi}_{xxx}^{(2)} = 26 \pm 15$ pm/V.

Although a large second order nonlinearity can be induced in silicon, the pump and signal propagation constants ($k_\omega, k_{2\omega}$) are required to be phase-matched for efficient second harmonic generation, $2k_\omega - k_{2\omega} = 0$. This is not the case for the fundamental TE optical modes due to waveguide and modal dispersion, $k_\omega = 6.132\,\mu m^{-1}, k_{2\omega} = 16.627\,\mu m^{-1}$ at $\lambda_\omega$~2.29μm and $\lambda_{2\omega}$~1.145μm (see Methods-Numerical Simulation). Therefore, the pump and signal should be coupled only when both are in phase and decoupled when both are out of phase. This is referred to as quasi-phase matching. A spatially periodic electric field along the waveguide is required for quasi-phase matching pump and signal modes in silicon. Lateral junctions were placed with a period of $\Lambda = 1.44\,\mu m$ to realize the periodic electric field, shown in Figure 2.d. The period was selected to match two times the coherence length for first order quasi-phase matching [10, 23], $2k_\omega^{TE_{11}} - k_{2\omega}^{TE_{11}} + \frac{2\pi}{\Lambda} = 0$. The generated second harmonic power $P_{2\omega}$ for a quasi-phase matched nonlinear media was derived using nonlinear coupled mode theory and the undepleted-pump approximation in [10],

$$P_{2\omega} = \frac{8\bar{\chi}_{xxx}^{(2)} L_{qpm}^2 P_\omega^2}{\varepsilon_0 n_\omega^2 n_{2\omega} c \lambda_\omega^2 A} \exp\left[-(2\alpha_\omega + \alpha_{2\omega})L/2\right] \frac{\sin^2(\Delta k L_{qpm}/2) + \sinh^2\left[(2\alpha_\omega - \alpha_{2\omega})L/4\right]}{(\Delta k L_{qpm}/2)^2 + \left[(2\alpha_\omega - \alpha_{2\omega})L/4\right]^2} \quad (2)$$

where $n_\omega$=2.245 and $n_{2\omega}$=3.043 are the effective refractive indices at the pump and signal wavelengths, $\alpha_\omega$=3.6cm$^{-1}$ and $\alpha_{2\omega}$=0.2cm$^{-1}$ are the simulated optical power loss coefficients at pump and signal wavelengths, $P_\omega$ is the pump power, $A$=0.0915μm$^2$ is the modal area, $L$=1mm and $L_{qpm}$=0.5mm are the nonlinear waveguide and the quasi-phase matched section lengths, $\Delta k = k_{2\omega}^{TE_{11}} - 2k_{\omega}^{TE_{11}} - 2\pi/\Lambda = 2\pi n_{2\omega}/\lambda_{2\omega} - 4\pi n_{\omega}/\lambda_{\omega} - 2\pi/\Lambda$ is the phase mismatch between the pump and second harmonic signal including the quasi-phase matched period. The pump power dependence of the SHG was calculated using Eq. (2), shown in Figure 3.a. The SHG efficiency was estimated to be within $0.9\%/W \leq P_{2\omega}/P_{\omega}^2 \leq 12\%/W$. The spectral response of SHG at $P_\omega$=25mW was also calculated using Eq. (2), shown in Figure 3.b. The spectral bandwidth of the main lobe was 6.5nm. Furthermore, when the waveguide width and the spatial period are altered in different waveguides, the quasi-phase matched pump wavelength was designed to be within $\lambda_\omega = 2\lambda_{2\omega} = 2.15\mu m$ and $\lambda_\omega = 2\lambda_{2\omega} = 2.42\mu m$.

The designed second harmonic generators were fabricated on a 300mm SOI wafer (Figure 3.c). A near-infrared CW tunable pump laser was free-space coupled to one end of a single mode fiber (SMF-2000) and the other end of the fiber was cleaved. The cleaved fiber end was used to couple pump laser to an on-chip inverse silicon taper. A polarization controller was used to align the linearly polarized output of the SMF-2000 and the fundamental TE mode of the on-chip waveguide. The on-chip pump laser power was calibrated using waveguides with varying lengths. The maximum on-chip pump power around $\lambda_\omega \sim 2.29\mu m$ was measured to be $P_\omega$=25mW (Figure 4.a), limited by the coupling losses and pump laser. The second harmonic signal at $\lambda_\omega \sim 1.145\mu m$ was collected using a lensed single mode fiber (SMF-28) and the fiber-to-chip coupling losses were calibrated using waveguides with varying lengths. The lateral p-i-n junctions within the waveguide were DC biased spanning from 0.5V to -21V. SHG was not observed when the diode was forward biased ($V_{dc}$>0.5V), indicating negligible background SHG from stress and silicon-SiO$_2$ interface contributions. The second harmonic signal was recorded as a function of the applied DC bias, shown in Figure 4.b. The maximum SHG signal was measured at $V_{DC}$=-21V. The reverse bias current passing through the p-i-n junction was below 0.1μA at $V_{DC}$=-21V and electric fields were below breakdown field. The maximum SHG efficiency was $P_{2\omega}/P\omega^2$=12%/W at a pump wavelength of $\lambda_\omega$~2.29μm. When the measurement results were overlaid with the simulation results in Figure 3.a, the effective field-induced second order nonlinear susceptibility was determined to be $\bar{\chi}^{(2)}_{xxx}$ = 41pm/V. The SHG spectrum was also measured as a function of reverse bias voltage, and showed the expected sinc$^2$-like response (Figure 4.c). The spectral response at $V_{DC}$=-21V is overlaid with the simulations in Figure 3.b. Furthermore, when the waveguides with different widths and spatial periods were tested, SHG was observed at multiple wavelengths spanning from $\lambda_\omega=2\lambda_{2\omega}$=2.16μm to $\lambda_\omega=2\lambda_{2\omega}$=2.42μm, agreeing well with the simulated quasi-phase matched pump wavelengths (Figure 4.d).

In conclusion, we have demonstrated field-induced second order nonlinear susceptibility in silicon waveguides using CMOS compatible fabrication methods. The origin of this second order nonlinearity is the large third order nonlinear susceptibility of silicon combined with large electric fields generated by p-i-n junctions, breaking the crystalline symmetry of silicon. The DC Kerr effect in silicon is used as an optical phase shifter in a MZI, demonstrating a $V_\pi L$ of 2.4Vcm. A quasi-phase matched EFISHG is demonstrated with a conversion efficiency of $P_{2\omega}/P\omega^2$=12%/W at $\lambda_\omega=2\lambda_{2\omega}$=2.29μm in a 1mm long ridge silicon waveguide. This corresponds to an effective field-induced $\bar{\chi}^{(2)}$ of 41pm/V. When the waveguide width and the spatial period was changed, SHG was measured at multiple wavelengths spanning from $\lambda_\omega$=2.16μm to $\lambda_\omega$=2.42μm. The efficiency can be further increased using a longer silicon waveguide. The spectral bandwidth can be also improved by chirping the quasi-phase matching period. Furthermore, the field-induced $\chi^{(2)}$ in silicon can be used in sum and difference frequency generation and electro-optic modulation.

**Methods**

**Device Fabrication**

The silicon ridge waveguide with an integrated lateral junction was fabricated in a 300mm CMOS foundry using silicon-on-insulator (SOI) wafers with a 225nm top silicon layer and a 2μm buried oxide (BOX) layer for optical isolation. The SOI was fully etched to form the waveguides. It was followed by a partial silicon etch to form the ridge waveguides. The etch depth was 110nm. An oxidization step was used to passivate the sidewalls, which reduced the full waveguide thickness to 220nm and the ridge thickness to 100nm. The p-i-n junction was formed by arsenic (As) and boron difluoride ($BF_2$) implants with target doping concentrations of $10^{18}$/cm$^3$. The $n+$ and $p+$ doped contact regions are formed by phosphorus and $BF_2$ implants with target concentrations of ~$10^{20}$/cm$^3$. The tungsten vias are contacted to highly doped regions by self-aligned silicidation. Two copper routing layers are used to contact to on-chip ground-signal-ground (GSG) direct current (DC) probing pads (60μm×60μm) at a 100μm pitch.

**Numerical Simulations**

The silicon and $SiO_2$ refractive indices for pump and signal wavelengths were determined using the fit parameters to the Sellmeier's equation in [24]. The carrier distribution within the silicon ridge waveguide was simulated using Synopysy's Sentaurus suite. The carrier distribution was converted to electro-refractive index distribution using the fit parameters for plasma-dispersion effect in [15]. These fit parameters were extracted at $\lambda_{er}$~1.55μm. The electro-refractive index distributions at pump and signal wavelength were scaled by $(\lambda_\omega/\lambda_{er})^2$ and $(\lambda_{2\omega}/\lambda_{er})^2$, following the plasma-dispersion relation in [11]. The refractive indices and electro-refractive index distributions at pump and signal wavelengths were combined, respectively. The mode profiles and complex propagation constants for the resulting index distributions were simulated using a finite-difference modesolver [13]. The loss coefficients at pump and signal wavelengths were extracted from the imaginary part of the complex propagation constants.

**Acknowledgements**

This work was supported by the Defense Advanced Research Projects Agency (DARPA) Microsystems Technology Office's (MTO) E-PHI (HR0011-12-2-0007) and DODOS (HR0011-15-C-0056) projects. Authors thank for the support of program managers, Josh Conway and Robert Lutwak.

**Author Contributions**

E. T. and M.R.W. conceived the idea of the project. E. T. simulated and designed nonlinear silicon waveguides, laid out the mask, conducted experiments on second harmonic generators, analyzed the results and wrote the manuscript. C. V. P. conducted experiments on DC Kerr mach-zehnder interferometers and edited the manuscript. M. R. W. also edited the manuscript and supervised the project.

**Additional Information**

**Competing financial interests :** The authors declare no competing financial interests.

**Figures**

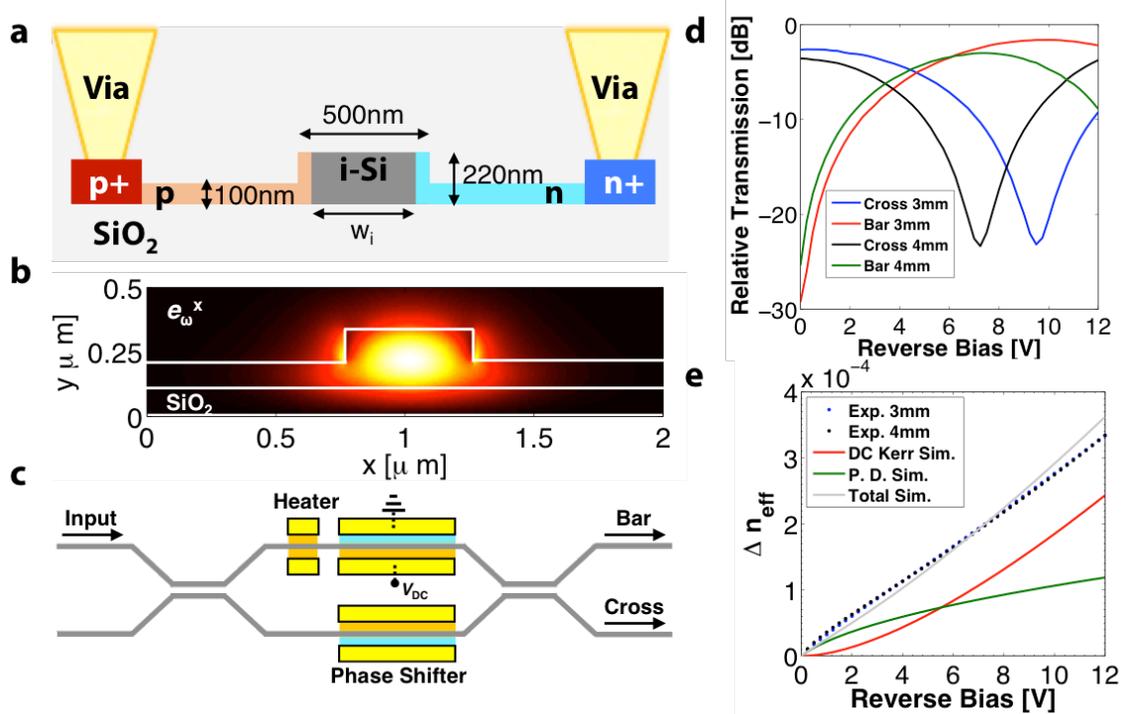

**Figure 1 | The DC Kerr effect in silicon. (a)** The p-i-n junction inside the ridge silicon waveguide and vias are sketched to indicate important device dimensions. (b) The modal electric field profile ($e_\omega^x$) for fundamental TE excitation is illustrated. (c) The ridge waveguide is used as a phase shifter in a Mach-Zehnder interferometer (MZI). **(d)** The measured transmission spectra of the bar and cross ports of MZIs with 3mm and 4mm long phase shifters are plotted as a function of DC bias. **(e)** The measured refractive index perturbations are plotted as a function of DC bias for both MZIs. The simulated DC Kerr, plasma dispersion (P. D.) and total index perturbations are overlaid with the experimental data for comparison.

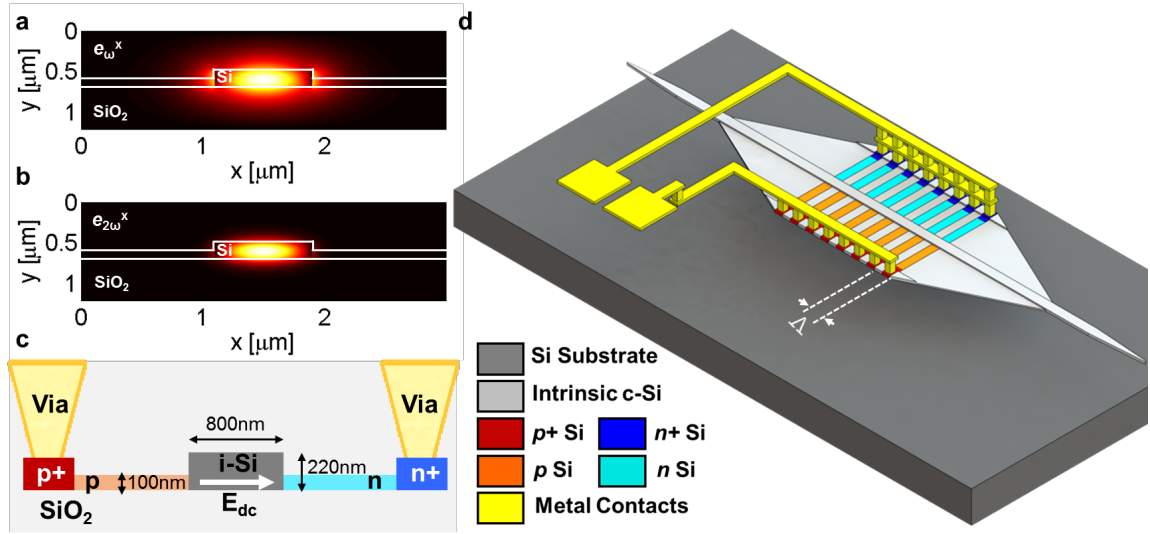

**Figure 2 | The design of a field-induced second harmonic generator. (a-b)** The modal electric field profiles ($e_\omega^x, e_{2\omega}^x$) at pump and second harmonic signal wavelengths are illustrated, showing maximal confinement inside the silicon core and a large overlap. **(c)** The p-i-n junction inside the ridge silicon waveguide and vias are sketched to indicate important device dimensions. **(d)** 3-D sketch of the ridge silicon waveguide and quasi-phase matched spatially periodic layout of the p-i-n junction. The dimensions are not to scale.

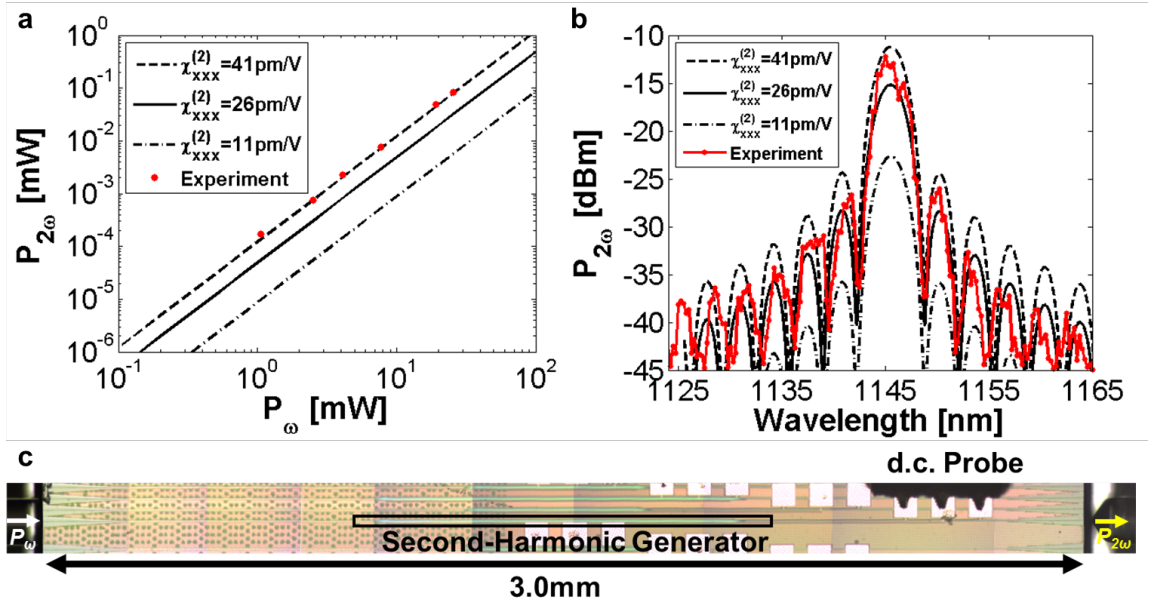

**Figure 3 | Analytical and experimental SHG. (a-b)** The analytically calculated on-chip second harmonic generation curves for estimated maximum, average and minimum second order susceptibilities are plotted as a function of pump power and wavelength. The measurement data is overlaid for comparison, showing an effective second order nonlinearity of $\bar{\chi}^{(2)}_{xxx} = 41\text{pm/V}$. The p-i-n junction was reverse biased at 21V, ($E^{x}_{DC} \sim 25 V\mu m^{-1}$). The CW pump power was $P_\omega$=25mW in (b). **(c)** The microscope image of the integrated second harmonic generator is shown with the single mode fibers (left: SMF-2000, right: SMF-28) and the DC ground-signal-ground probe.

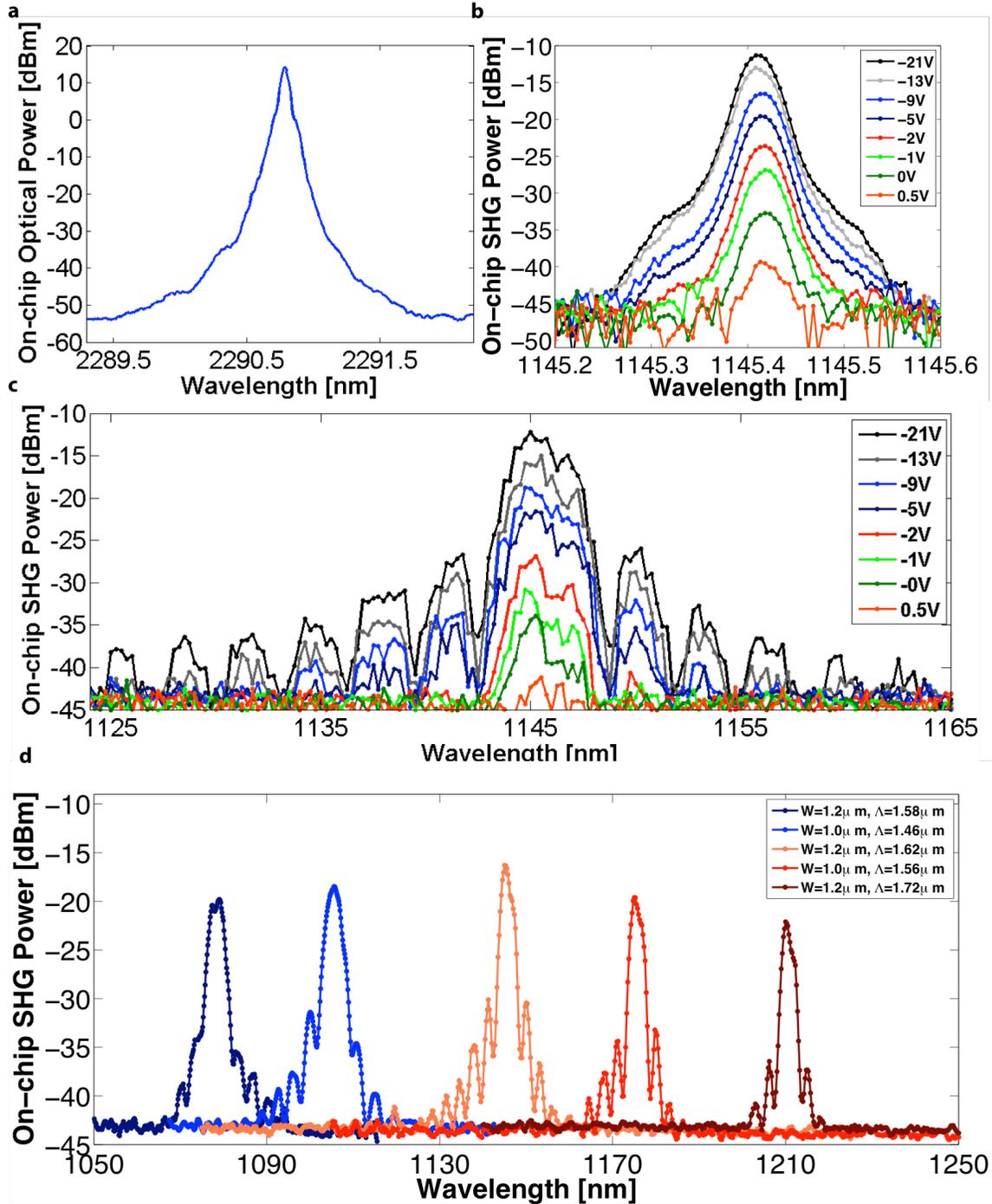

**Figure 4 | Voltage dependence of the SHG. (a-b)** The on-chip CW optical pump and generated second harmonic powers are measured as a function of wavelength and applied voltage bias using a spectrum analyzer. The CW pump power was $P_\omega$=25mW. The SHG power got stronger with a larger reverse bias voltage, following EFISHG theory. **(c)** The spectral dependence of the SHG power was measured as a function of wavelength and applied voltage bias using a spectrum analyzer. (d) The SHG of multiple waveguides with different quasi-phase matching periods are plotted for different wavelengths. The CW pump power was $P_\omega$=19mW and $E_{DC}^x \sim 25 V \mu m^{-1}$ at all wavelengths.